\documentclass{ws-mpla}

\usepackage{amsmath}
\usepackage{amssymb}

\newcommand{\nslash}{n\!\!\! /}
\newcommand{\bnslash}{\bar n\!\!\! /}
\newcommand{\Aslash}{A\!\!\! /}
\newcommand{\Dslash}{D\!\!\!\!/\,}
\newcommand{\Pslash}{\mathcal{P}\!\!\!\!/}
\newcommand{\vect}[1]{\mathbf{#1}}
\newcommand{\abs}[1]{\left\lvert #1\right\rvert}
\newcommand{\bra}[1]{\left\langle #1\right\rvert}
\newcommand{\ket}[1]{\left\lvert #1\right\rangle}
\newcommand{\Lqcd}{\Lambda_{\text{QCD}}}
\newcommand{\e}{\mathrm{e}}
\newcommand{\Cumulant}[1]{\left\langle\!\!\!\left\langle #1 \right\rangle\!\!\!\right\rangle}
\newcommand{\cumulant}[1]{\langle\!\langle #1 \rangle\!\rangle}
\newcommand{\SCETa}{$\text{SCET}_{\text{I}}$}
\newcommand{\SCETb}{$\text{SCET}_{\text{II}}$}

\DeclareMathOperator{\Tr}{Tr}

\begin{document}

\markboth{Christopher Lee}
{Nonpertubative Effects in Event Shapes from SCET}

%%%%%%%%%%%%%%%%%%%%% Publisher's Area please ignore %%%%%%%%%%%%%%
\catchline{}{}{}{}{}
%%%%%%%%%%%%%%%%%%%%%%%%%%%%%%%%%%%%%%%%%%%%%%%%%%%%%%%%%%%%%%%%%%%

\title{UNIVERSAL NONPERTURBATIVE EFFECTS IN EVENT SHAPES FROM SOFT-COLLINEAR EFFECTIVE THEORY}

\author{\footnotesize CHRISTOPHER LEE}

\address{Institute for Nuclear Theory, University of Washington, Seattle, WA 98195-1550, USA\\
clee@phys.washington.edu}

\maketitle

\pub{Received 1 February 2007}{ }

\begin{abstract}
Two-jet event shape distributions, traditionally studied in the language of perturbative QCD, can be described naturally in soft-collinear effective theory. In this language, we demonstrate factorization of event shape distributions into perturbatively-calculable hard and jet functions and nonperturbative soft functions, and show how the latter contribute universal shifts to the mean values of various event shape distributions. Violations of universality in shifts of higher moments can give information on correlations of energy flow in  soft radiation.

\end{abstract}

\ccode{PACS Nos.: 12.39.St, 13.66.Bc, 13.87.-a}

\section{Introduction}

Hadronic jets produced in $e^+e^-$ annihilations provide a rich laboratory for probing both the perturbative and nonperturbative behavior of strong interactions described by Quantum Chromodynamics (QCD). Studying these leptonic collisions allows us to focus on strong effects only in the final state. Asymptotic freedom makes possible the reliable description in perturbation theory of processes at high center-of-mass energies $Q$, such as the prediction of dominance of two- and three-jet-like states in the hadronic final states in $e^+e^-$ annihilation and their respective cross-sections.\cite{StermanWeinberg} The nonperturbative process of hadronization at the low energy scale $\Lqcd$, however, introduces corrections which we do not yet know how to calculate. For some observables, such as the total hadronic cross-section, these corrections are small, of order $(\Lqcd/Q)^4$.\cite{Peskin} In less inclusive observables, these corrections are larger, on the one hand increasing uncertainty due to nonperturbative corrections, but, on the other, giving us an opportunity to measure their size and probe some of their properties even if we cannot calculate them exactly.

One such class of observables is that of \emph{event shapes}, numbers that depend on the distrbution of momenta of final state hadrons, probing in greater detail the structure of the jets than does the total hadronic or $n$-jet cross section. The study of $e^+e^-$ event shapes has a long history in the framework of perturbative QCD (pQCD).\cite{review} A two-jet event shape $e$ is defined to range between 0 and 1, with exactly back-to-back pencil-like jets corresponding to $e=0$. Predicting the distribution of events near the region $e=0$ requires both perturbative resummation as well as an accounting of nonperturbative power corrections, which enter as powers of $\Lqcd/(eQ)$ due to nonperturbative soft or collinear radiation from jets.

Many approaches have led to fruitful analysis of these power corrections, mostly based on the behavior of perturbation theory predictions for event shape distributions. Renormalon ambiguities of order $\Lqcd/Q$ in the perturbative series for jet observables led to the prediction of compensating $\Lqcd/Q$ nonperturbative power corrections.\cite{MW94,oneoverQ1,oneoverQ2,oneoverQ3,oneoverQ4,oneoverQ5,oneoverQ6,oneoverQ7,KS95,DW1,DMW} These studies led to models of a universal effective or dispersive infrared coupling replacing the strong coupling at low scales implying universal power corrections to different event shapes,\cite{DW1,DMW} and later to the model of \emph{dressed gluon exponentiation} (DGE),\cite{Gardi1,Gardi2,GardiFRIF} combining the resummation of perturbative logarithms and nonperturbative corrections from renormalons.   Section 2 introduces the event shapes we choose to study and reviews the predictions of universality. 

Another powerful tool in the study of power corrections is factorization,  separating  perturbative and nonperturbative effects in jet observables.\cite{KS95,pqcdfact1,pqcdfact2,KS99} Event shape distributions can be written as convolutions of perturbatively-calculable and nonperturbative functions, written schematically:
\begin{equation}
\frac{d\sigma}{de} = H\otimes J_1\otimes J_2\otimes S,
\label{factorizationtheorem}
\end{equation}
where $H$ describes the hard scattering, $J_{1,2}$ describe the dynamics within the jets, and $S$ describes soft radiation away from the jets. The soft function $S$ can be viewed as a resummation of all powers of $\Lqcd/(eQ)$ in the region $e\sim\Lqcd/Q$, but we can also look over a larger range of $\Delta e$, and extract information about the low-order terms in $\Lqcd/(eQ)$ from $S$. 

The separation of scales (hard, jet, and soft) that allows a factorization of the form of Eq.~(\ref{factorizationtheorem}) suggests the formulation and use of an effective field theory description in calculating the event shape distributions. The \emph{soft-collinear effective theory}\cite{SCET1,SCET2}  (SCET) is the theory appropriate for this task. Originally applied mostly to problems in $B$ physics, SCET also provides a natural description for the environment of jet physics, and has been used not only to analyze nonperturbative power corrections\cite{BMW,BLMW,A001,LS07} but also the improvement of event generators\cite{BS1,BS2} and the perturbative calculation of jet cross sections\cite{Trott}.  In this review we focus on the identification of the nonperturbative power corrections, and not so much on the perturbative issues of matching, running, and resummation. Section 3 provides a brief overview of aspects of the SCET formalism that we need in this discussion.

In Sec.~4, we use SCET to reproduce the factorization theorem (\ref{factorizationtheorem}) for event shape distributions\cite{BMW,BLMW}, already well-known from arguments in pQCD\cite{KS95,KS99,K98}, and identify more precisely what is meant by the schematic notation in (\ref{factorizationtheorem}). We will not be able to provide a detailed description of the pQCD treatment of event shapes in this review, focusing instead on the SCET formulation. However, a comparison of the treatments in SCET and full QCD has not only elucidated tight connections between the two approaches, but has also inspired proofs of the universal relations among power corrections to different event shapes which go beyond some of the assumptions implicit in the pQCD-based models.\cite{A001,LS07} We review these developments in Sec.~5, and conclude in Sec.~6.

\section{Event Shapes}

\subsection{Two-Jet Event Shapes}

The event shapes we shall consider are defined as functions of the momenta $p_i$ of the particles in the hadronic final state $N$ in the collisions $e^+e^-\rightarrow N$. They are ``traditional'' event shapes such as the \emph{thrust}:
\begin{equation}
T = \frac{1}{Q}\max_{\hat{\vect{t}}}\sum_{i\in N}\abs{\hat{\vect{t}}\cdot\vect{p}_i},
\end{equation}
the \emph{jet broadening}:
\begin{equation}
B = \frac{1}{Q}\sum_{i\in N}\abs{\hat{\vect{t}}\times\vect{p}_i},
\end{equation}
and the \emph{$C$ parameter}:
\begin{equation}
C = \frac{3}{2Q^2}\sum_{i\in N}\abs{\vect{p}_i}\abs{\vect{p}_j}\sin^2\theta_{ij};
\end{equation}
and the more recently introduced \emph{angularities}\cite{BKS03}:
\begin{equation}
\tau_a = \frac{1}{Q}\sum_{i\in N}E_i\sin^a\theta_i(1-\cos\theta_i)^{1-a},
\end{equation}
where the angles $\theta_i$ are measured with respect to the thrust axis. The angularity $\tau_a$ is an infrared-safe observable for $a<2$, although our analysis of nonperturbative power corrections below will be valid only for $a<1$. Between $a=0$ and $a=1$, the angularities interpolate between the thrust ($\tau_0 = 1-T$) and the jet broadening ($\tau_1 = B$). Adjusting the continuous parameter $a$ to move through this family of similar event shapes will allow us to extract more information about nonperturbative physics from the power corrections than would be possible by studying individual event shapes separately.

These event shapes can be rewritten in a common form, using the rapidities $\eta_i$ of final state particles, where $\eta_i = \frac{1}{2}\ln\left(\frac{E_i + p_i^z}{E_i - p_i^z}\right)$, $z$ being along the thrust axis. For massless particles, $\eta_i = \ln\cot(\theta_i/2)$. The event shapes above take the general form:
\begin{equation}
e = \frac{1}{Q}\sum_{i\in N}\abs{\vect{p}_i^\perp}f_e(\eta_i),
\label{general}
\end{equation}
where the appropriate choice of the function $f_e$ yields the various event shapes. For the angularities, $f_{\tau_a}(\tau_a)  = \exp[-\abs{\eta_i}(1-a)]$, and for the $C$-parameter, $f_C = 3/\cosh\eta_i$. The form (\ref{general}) will be useful for studying the behavior of soft gluon radiation under Lorentz boosts along the thrust axis.

\subsection{Power Corrections from Perturbative QCD}

Analyses based on perturbative QCD predict a simple behavior for the leading power corrections to event shape distributions due to soft radiation. Many of these are based on finding the shift in the resummed perturbative event shape distributions due to the (exponentiated) emission of a single gluon with small transverse momentum, $k^\perp\sim\Lqcd$. The consistent prediction of such approaches is a universal power correction shifting the perturbative predictions of the mean values\cite{DW1} or even the entire distribution\cite{DW2} of event shapes:
\begin{equation}
\langle e\rangle = \langle e\rangle_{\text{PT}} + c_e\frac{\mathcal{A}}{Q},\qquad\frac{d\sigma}{de}(e)= \frac{d\sigma}{de}\biggr\rvert_{\text{PT}}\left(e - c_e\frac{\mathcal{A}}{Q}\right) ,
\label{universalshift}
\end{equation}
where $c_e$ is an observable-dependent but exactly-calculable coefficient and $\mathcal{A}$ is an unknown but universal nonperturbative parameter. This universal structure persists up to $\mathcal{O}(\alpha_s^2)$ in soft gluon emission, with the single-gluon shift, $c_e\mathcal{A}/Q$, just  multiplied by another universal number, the \emph{Milan factor} $\mathcal{M}$.\cite{Milan}  In approaches modeling the strong coupling with an effective infrared\cite{DW1} or dispersive coupling\cite{DMW} in the low-momentum regime, $\mathcal{A}$ depends on the IR coupling $\bar\alpha_0(\mu_I)$, where $\mu_I$ is the infrared-cutoff scale dividing perturbative and nonperturbative regimes, and the perturbative strong coupling $\alpha_s(\mu)$ at a higher scale. Fits to the shift in mean values of various event shapes observed in data yield a value of $\alpha_s(M_Z)$ consistent with other extractions and a universal value of $\bar\alpha_0(\mu_I)$ at $\mu_I = 2\text{ GeV}$ of about 0.50.\cite{review} Good fits to the event shape distributions over the whole two-jet region and a wide range of $Q^2$ have been obtained using the DGE cross-sections.\cite{Gardi1,Gardi2}

Similar arguments applied to the angularity distributions predict a remarkably simple scaling behavior for the entire soft function appearing in the factorized angularity distributions.\cite{scaling1,scaling2,Magnea06} The distribution in $\tau_a$ takes the factorized form:
\begin{equation}
\frac{d\sigma}{d\tau_a}(\tau_a) = \int d\xi \sigma_J^a(\tau_a - \xi)S_a(\xi),
\end{equation}
where $\sigma_J^a$ is the perturbatively-calculated $\tau_a$ distribution, and $S_a(\xi)$ is the soft function. Analyzing the contribution of low-momentum gluon radiation in the NLL resummed cross-section, the soft function is shown to exponentiate,
\begin{equation}
\tilde S_a(\nu) = \exp\left[\frac{1}{1-a}\sum_{n=1}^\infty\lambda_n\left(-\frac{\nu}{Q}\right)^n\right],
\label{angularityexponentiated}
\end{equation}
transformed to moment space by $\tilde S_a(\nu) = \int_0^\infty d\xi \e^{-\nu\xi}S_a(\xi)$.  According to this prediction, the coefficients $\lambda_n$ are independent of $a$, and the various angularity shape functions are related to the thrust shape function by:
\begin{equation}
\tilde S_a(\nu) = \tilde S_0(\nu)^{\frac{1}{1-a}}.
\label{scalingrule}
\end{equation}
This relation has been tested using simulated data from PYTHIA\cite{scaling1}, and awaits tests against data from LEP.

The predictions of these perturbative QCD-based models for the soft power corrections are quite successful, but as they rely on the behavior of perturbation theory for low-order gluon emission, seem not to capture fully the nonperturbative effects to all orders of soft gluon emission. To move beyond this limitation, we will find useful an approach that clearly separates the perturbative and nonperturbative physics, provided for us by the tools of effective field theory.

\section{Soft-Collinear Effective Theory}

In this section we provide a brief but mostly self-contained overview of the formalism of soft-collinear effective theory.

\subsection{Soft and Collinear Modes}

In a two-jet event, there are several distinct momentum scales: the total center-of-mass energy $Q$, the scale of nonperturbative radiation $\Lqcd$, and the intermediate scale $\sqrt{Q\Lqcd}$ describing the typical transverse momentum of particles in the jets. The existence of this hierarchy of scales leads to the construction of an effective field theory, the \emph{soft-collinear effective theory} as an expansion in the parameter $\lambda = \sqrt{\Lqcd/Q}$.\cite{SCET1,SCET2} The modes in this effective theory characterized by the scaling of the light-cone components of their momenta, $p = (n\cdot p,\bar n\cdot p,p^\perp)$, with respect to back-to-back lightlike vectors $n,\bar n$. \emph{Collinear} momenta scale as $p_c\sim Q(\lambda^2,1,\lambda)$, and \emph{ultrasoft} (usoft) momenta scale as $p_{us} \sim Q(\lambda^2,\lambda^2,\lambda^2)$. The theory with this collinear scaling is called \SCETa, and is appropriate to describe jets whose constituents have typical transverse momenta of the order $p_c^\perp\sim \sqrt{Q\Lqcd}$. Narrower jets may be described in a theory \SCETb\ in which collinear momenta scale as $p_c\sim Q(\lambda^4,1,\lambda^2)$.\cite{SCETII} We will study event shapes dominated by jets wide enough to be treated in \SCETa, and the shortened abbreviation ``SCET'' will always refer to \SCETa\ in the remainder of this review.

The fields in the SCET Lagrangian are collinear quarks and gluons, $\xi_n$ and $A_n$, and ultrasoft quarks and gluons, $q_{us}$ and $A_{us}$. For the collinear modes, we split their momenta into two parts, a large \emph{label} piece and a \emph{residual} piece: $p_n = \tilde p_n + k$, where
\begin{equation}
\tilde p_n^\mu = \bar n\cdot \tilde p_n\frac{n^\mu}{2} + \tilde p_n^\perp
\end{equation}
contains only the $\mathcal{O}(Q)$ piece of $\bar n\cdot p_n$ and the $\mathcal{O}(Q\lambda)$ piece of $p_n^\perp$, leaving the residual momentum $k$ of order $Q\lambda^2$ in all components. The collinear and soft fields are defined by factoring out a phase removing the dependence on the label momenta from the full theory fields $q$ and $A$:
\begin{equation}
q(x) = \sum_{\tilde p\not=0}\e^{-i\tilde p\cdot x}q_{n,p}(x) + q_{us}(x),\qquad A(x) = \sum_{\tilde p\not =0}e^{-i\tilde p\cdot x}A_{n,p}(x) + A_{us}(x). 
\label{QCDfields}
\end{equation}
The fields $q_{n,p}(x)$ and $A_{n,p}(x)$ now have fluctuations in spacetime described by momenta of order $Q\lambda^2 = \Lqcd$ about the corresponding label momentum $\tilde p$. The restriction of the sums to nonzero labels is crucial, to avoid double-counting the momenta already covered by the usoft modes $q_{us}$ and $A_{us}$.\cite{0bin} It is useful to define operators\cite{labelops} which pick out only the label components of the momentum of a collinear field $\phi_{n,p}$:
\begin{equation}
\mathcal{P}_\mu\phi_{n,p} = \tilde p_\mu\phi_{n,p},\quad\mathcal{P}_\mu\phi_{n,p}^* = -\tilde p_\mu\phi_{n,p}^*.
\end{equation}
We will also use the shorthand $\bar{\mathcal{P}} \equiv \bar n\cdot \mathcal{P}$, and $\mathcal{P}_\perp$.

\subsection{The SCET Lagrangian}

To complete the construction of the SCET Lagrangian, the components 
\begin{equation}
\xi_{n,p} = \frac{\nslash\bnslash}{4}q_{n,p},\qquad \Xi_n=\frac{\bnslash\nslash}{4}q_{n,p}
\end{equation}
are projected out of the field $q_{n,p}$. Writing the QCD Lagrangian in terms of these fields, it can be shown that $\xi_{n,p}$ is massless while $\Xi_{n,p}$ has an effective mass $\bar n\cdot\tilde p \sim\mathcal{O}(Q)$, which is thus integrated out of the theory. Keeping only the terms of leading power in $\lambda$, we obtain the leading-order Lagrangian of SCET:
\begin{equation}
\mathcal{L}_{\text{SCET}} = \mathcal{L}_{cq} + \mathcal{L}_{cg} + \mathcal{L}_{us},
\end{equation}
where the collinear quark Lagrangian $\mathcal{L}_{cq}$ is\cite{SCET2}
\begin{equation}
\mathcal{L}_{cq} = \bar\xi_{n,p'}\left[in\cdot D_{us} + gn\cdot A_{n,q} + (\Pslash_\perp + g\Aslash^\perp_{n,q})W_n\frac{1}{\mathcal{\bar P}}W_n^\dag(\Pslash_\perp + g\Aslash^\perp_{n,q'})\right]\frac{\bnslash}{2}\xi_{n,p},
\end{equation}
where $W_n(x) = P\exp[ig\int_{-\infty}^0ds\,\bar n\cdot A_n(\bar ns+x)]$ is a Wilson line of collinear gluons;
the collinear gluon Lagrangian is\cite{redef}
\begin{equation}
\mathcal{L}_{cg} = \frac{1}{2g^2}\Tr\left\{[i\mathcal{D}^\mu+gA^\mu_{n,q},i\mathcal{D}^\nu+gA^\nu_{n,q'}]^2\right\} + \mathcal{L}_{c}^{g.f.},
\end{equation}
where $i\mathcal D^\mu = \mathcal P^\mu + \frac{\bar n^\mu}{2}in\cdot D_{us}$ and $\mathcal{L}_c^{g.f.}$ contains the collinear ghost and gauge-fixing terms;
and the ultrasoft Lagrangian,
\begin{equation}
\mathcal{L}_{us} = \bar q_{us}i\Dslash_{us}q_{us} - \frac{1}{2}\Tr G_{\mu\nu}G^{\mu\nu} + \mathcal{L}_{us}^{g.f.},
\end{equation}
is simply that of QCD. The ultrasoft covariant derivative $iD_{us}^\mu = i\partial^\mu + gA_{us}^\mu$ contains only the ultrasoft gauge field, the ultrasoft gauge field strength is $G^{\mu\nu} = (i/g)[D_{us}^\mu,D_{us}^\nu]$, and $\mathcal{L}_{us}^{g.f.}$ contains the ultrasoft gauge fixing and ghost terms. In the collinear Lagrangians, the label momentum phase factors and sums over labels are implicit. These simply have the effect of enforcing label momentum conservation in each term.\cite{labelops}

\subsection{Decoupling Soft and Collinear Modes}

In the leading-order SCET Lagrangian, usoft quarks do not interact with collinear modes, and the coupling of usoft gluons to collinear modes is particularly simple, with only the $n\cdot A_{us}$ component appearing. These interactions can be made to ``disappear'' from the collinear Lagrangians by the following field redefinitions:\cite{redef}
\begin{equation}
\xi_n = Y_n^\dag\xi'_n,\quad A_n = Y_n^\dag A_n' Y_n,
\label{Yredef}
\end{equation}
using the outgoing\footnote{The field redefinition may be performed with incoming Wilson lines, with $s = -\infty$ to 0, depending on whether the collinear particles to be described are incoming or outgoing in the hard scattering or decay.\cite{Chay}} Wilson line of ultrasoft gluons,
\begin{equation}
Y_n(z) = P\exp\left[ig\int_0^\infty ds\,n\cdot A_{us}(ns +z)\right].
\end{equation}
The redefinition (\ref{Yredef}) also implies $W_n = Y_n^\dag W_n' Y_n$. Since $in\cdot D_{us} Y_n^\dag = 0$, making these substitutions for the collinear fields in $\mathcal{L}_{cq}$ and $\mathcal{L}_{cg}$ removes their couplings with usoft gluons:
\begin{gather}
\bar\xi_{n,p'} in\cdot D_{us}\xi_{n,p} = \bar\xi_{n,p'}' in\cdot\partial \xi_{n,p}', \\
\Tr\left\{[i\mathcal{D}^\mu+gA^\mu_{n,q},i\mathcal{D}^\nu+gA^\nu_{n,q'}]^2\right\}  = \Tr\left\{[i\mathcal{D}_{(0)}^\mu+g{A'}^\mu_{n,q},i\mathcal{D}_{(0)}^\nu+g{A'}^\nu_{n,q'}]^2\right\} ,
\end{gather}
where $i\mathcal{D}_{(0)}^\mu = \mathcal{P}^\mu + \frac{\bar n^\mu}{2}in\cdot\partial$. This absence of interactions between the collinear and ultrasoft sectors simplifies proofs of factorization at leading order in the SCET expansion. Effective theory operators built out of collinear fields will always have ultrasoft Wilson lines accompanying them.

\section{Factorization of Two-Jet Event Shape Distributions}

\subsection{Matching the Two-Jet Current}

The two-jet event shape distributions in $e^+e^-\rightarrow\gamma^*\rightarrow N$ are calculated from matrix elements of the current $j^\mu = \bar q\gamma^\mu q$ in QCD:
\begin{equation}
\frac{d\sigma}{de} = \frac{1}{2Q^2} \sum_N\abs{\bra{N}j^\mu(0)\ket{0} L_\mu}^2 (2\pi)^4\delta^4(Q - p_N)\delta(e - e(N)),
\label{QCDdistribution}
\end{equation}
where $L_\mu$ is the leptonic part of the amplitude. We want this current to create to create states with two hadronic jets back-to-back in the center-of-mass frame. So, we will match this current in QCD onto operators in an SCET for collinear fields in two separate light-like directions\cite{hardscattering}, $n_1$ and $n_2$:
\begin{equation}
J^\mu_{n_1,n_2} = \bar\chi_{n_1,p_1}\gamma_\perp^\mu \chi_{ n_2,p_2},
\end{equation}
where we have introduced the jet field $\chi_n = W_n^\dag\xi_n$. The presence of the collinear Wilson line is required by gauge invariance.\cite{labelops}
The operators satisfy the matching condition:
\begin{equation}
\langle j^\mu\rangle_{\text{QCD}}(\mu) = C_{n_1 n_2}(\tilde p_1\cdot\tilde p_2;\mu)\langle J^\mu_{n_1,n_2}\rangle_{\text{SCET}}(\mu),
\end{equation}
where $C$ is the matching coeffcient, dependent on the matching scale $\mu$. A sum over $n_{1,2}$ is implicit. The matrix elements on both sides are taken between the same states, calculated in QCD and SCET, respectively. Here, our currents must produce a back-to-back $q\bar q$ pair in the final state, which requires $n_1 = \bar n_2$. To $\mathcal{O}(\alpha_s)$, the necessary matching coefficient is\cite{BMW,BS2}
\begin{equation}
C_{n\bar n}(\tilde p_1\cdot\tilde p_2;\mu) = 1 - \frac{\alpha_s(\mu)C_F}{4\pi}\left[8 - \frac{\pi^2}{6} + \ln^2\left(\frac{\mu^2}{\tilde p_1\cdot \tilde p_2}\right) + 3\ln\left(\frac{\mu^2}{\tilde p_1\cdot \tilde p_2}\right)\right]
\label{matching}
\end{equation}
The collinear labels must be $\tilde p_1 = Q n/2$ and $\tilde p_2 = -Q\bar n/2$ ($\xi_{n,\tilde p_2}$ creates an antiquark with label momentum $-\tilde p_2$)\cite{labelops}, so $\tilde p_1\cdot\tilde p_2 = -Q^2$. (For the current in deep inelastic scattering in the Breit frame, both labels are positive, and $\tilde p_1\cdot\tilde p_2 = Q^2$.)\cite{DIS} The matching coefficient simplifies at the scale $\mu=Q$:
\begin{equation}
C_{n\bar n}(-Q^2;Q) = 1 - \frac{\alpha_s(Q)C_F}{4\pi}\left(8 - \frac{7\pi^2}{6} + 3i\pi\right).
\end{equation}
The matching coefficient at another scale $\mu$ can be obtained via renormalization group evolution. The anomalous dimension of the two-jet operator is\cite{BS1,BS2}
\begin{equation}
\gamma_2(\mu) = -\frac{\alpha_s(\mu)C_F}{\pi}\left[\log\left(\frac{\mu^2}{-Q^2}\right) + \frac{3}{2}\right].
\end{equation} 
The accuracy of the effective theory prediction for the event shape distribution can be improved by including operators with higher numbers of jets\cite{BS1,BS2,Trott} (collinear fields in more directions).

\subsection{Factorizing the Distribution}

With this matching, the SCET expression for the event shape distribution becomes
\begin{equation}
\frac{d\sigma}{de} = \frac{\abs{C(\mu)}^2}{2Q^2}\sum_N\abs{\bra{N}T\bar\chi'_{n,Q} Y_n \gamma_\perp^\mu Y_{\bar n}^\dag \chi'_{\bar n,-Q}(0)\ket{0}L_\mu}^2(2\pi)^4\delta^4(Q-p_N) \delta(e - e(N)),
\label{SCETcrosssection}
\end{equation}
where we have made use of the collinear fields $\chi'_n = W'_n\xi'_n$ redefined according to Eq.~(\ref{Yredef}), so that the soft gluons appear in Wilson lines in the operator. The time-ordering which was implicit in the matrix element in Eq.~(\ref{QCDdistribution}) is written explicitly in Eq.~(\ref{SCETcrosssection}) since the Wilson lines contain fields at many different times.

To factorize the matrix element in (\ref{SCETcrosssection}) into purely collinear and soft parts, we can split the final state $\ket{N}$ into collinear and soft sectors, $\ket{N} = \ket{J_n}\otimes\ket{J_{\bar n}}\otimes\ket{X_u}$.\cite{BLMW,BKS03} Particles in the final state are assigned to these sectors according to their label momenta---particles with zero label momentum belong to the sector $X_u$.\cite{0bin} Since there are no interactions in the Lagrangian between the collinear fields $\xi'_{n,\bar n}, A'_{n,\bar n}$ and usoft gluons $A_{us}$, the matrix element in (\ref{SCETcrosssection}) factorizes to give:
\begin{align}
\frac{d\sigma}{de} = \frac{\abs{C}^2}{2Q^2}\sum_{J_n J_{\bar n}X_u}& \abs{\bra{J_n J_{\bar n}} T\bar\chi_{n,Q}'\gamma_\perp^\mu\chi'_{\bar n,-Q}(0)\ket{0}L_\mu}^2\frac{1}{N_C}\Tr\abs{\bra{X_u} T[Y_n (0) Y_{\bar n}^\dag(0)]\ket{0}}^2 \nonumber \\
&\times (2\pi)^4\delta^4(Q-p_{J_n}-p_{J_{\bar n}}-p_{X_u}) \delta(e - e(J_n) - e(J_{\bar n}) - e(X_u)),
\end{align}
where the trace is over colors.
We can write this factorized cross-section as a convolution of purely collinear and soft functions:
\begin{equation}
\frac{d\sigma}{de} = \int de_J \,\sigma_J(e_J)S_e(e-e_J),
\label{convolution}
\end{equation}
where $\sigma_J$ is the collinear cross-section
\begin{align}
\sigma_J(e_J) = \frac{\abs{C}^2}{2Q^2}\sum_{J_n J_{\bar n}}& \abs{\bra{J_n J_{\bar n}}\bar\chi_{n,Q}'\gamma_\perp^\mu\chi'_{\bar n,-Q}(0)\ket{0}L_\mu}^2 \nonumber \\
&\times(2\pi)^4\delta^4(Q-p_{J_n}-p_{J_{\bar n}})\delta(e_J - e(J_n) - e(J_{\bar n})),
\label{jetfunction}
\end{align}
which is equivalent to the perturbatively-calculated distribution $d\sigma/de\rvert_{\text{PT}}$,
and $S_e$ is the soft function
\begin{equation}
S_e(e) = \frac{1}{N_C}\Tr\sum_{X_u}\abs{\bra{X_u}T[Y_n Y_{\bar n}^\dag]\ket{0}}^2 \delta(e - e(X_u)).
\label{softfunction}
\end{equation}
To separate the two functions we neglected the contribution of the usoft momentum in the momentum-conserving delta function\cite{BKS03} in Eq.~(\ref{jetfunction}), which is a subleading contribution in $\Lqcd/Q$, and took the usoft momentum in the soft function (\ref{softfunction}) to be unrestricted. This is correct because the high momentum cutoff of the usoft integrals can be viewed as parametrically larger (in $1/\lambda$) than typical usoft momenta, so the upper limits are infinite in the limit $\lambda\rightarrow 0$.\cite{BLMW}

\subsection{Eliminating Double-Counting}

This procedure raises the question of double-counting. The soft function contains radiation emitted along the jets, overlapping with collinear radiation whose energy is small. This double-counting is avoided in SCET by always implementing the restriction to nonzero label momenta in the sums in (\ref{QCDfields}) for collinear modes when calculating collinear phase space or loop integrals. This procedure is known as \emph{zero-bin subtraction} and has important consequences in calculations in SCET.\cite{0bin} A related procedure in factorization theorems of perturbative QCD is the introduction of \emph{eikonal jet functions} and \emph{eikonal cross-sections} which are subtracted out of the na\"\i vely-defined jet and soft functions to eliminate double-counting.\cite{BKS03} Interpreting these functions in the language of SCET leads to a simple demonstration of the equivalence of the two subtraction methods, shown to leading order in $\lambda$ and all orders in $\alpha_s$ in Ref.~\cite{LS07}. Ref.~\cite{Idilbi} also illustrated this equivalence by explicit calculation at fixed orders in $\alpha_s$.

\subsection{Running the Jet and Soft Functions}

The collinear and jet functions in Eq.~(\ref{convolution}) depend on scale. As we saw in  Eq.~(\ref{matching}), the two-jet matching coefficient $C_{n\bar n}(\mu)$ contains logarithms of $\mu/Q$, and so is naturally evaluated at the scale $\mu=Q$. The collinear jet function and soft functions are naturally evaluated at scales $\mu_c\sim Q\lambda$ and $\mu_s\sim Q\lambda^2$. The running of the functions among these different scales requires knowing the anomalous dimensions of the collinear and soft matrix elements appearing in the jet and soft functions.\cite{Trott} For some event shapes, nonperturbative contributions to the collinear jet function at the scale $\mu_c\sim Q\lambda^2$ become important. In this case, we need to know how to run the collinear matrix element down to this scale. Between the scales $Q\lambda$ and $Q\lambda^2$, the scaling of the collinear momenta changes from that of \SCETa\ to that of \SCETb, and we must match the first theory onto the second. The fields and Lagrangian of \SCETb\ remain the same, only the scaling of the collinear momenta changes, and collinear matrix elements in \SCETb\ are nonperturbative.

We can estimate the size of nonperturbative power corrections in the collinear and soft functions for a particular event shape by considering the contribution of particles with transverse momenta of order $p^\perp\sim\Lqcd$ to these functions in \SCETb. Rapidities of collinear particles in \SCETb\ are of order $\abs{\eta}\sim\frac{1}{2}\abs{\ln(n\cdot p/\bar n\cdot p)}\sim\ln(1/\lambda^2)$, and those of soft particles are of order $\eta\sim 0$. In the case of angularities, the factor $f_{\tau_a}(\tau_a) = \e^{-\abs{\eta}(1-a)}$ for collinear particles is then of order $\lambda^{2(1-a)} = (\Lqcd/Q)^{1-a}$, and for soft particles, order 1. Then the nonperturbative collinear radiation contributes an amount of order $\abs{\vect{p}_\perp}\e^{-\abs{\eta}(1-a)}\sim (\Lqcd/Q)^{2-a}$, and the soft radiation an amount of order $\Lqcd/Q$, to the event shape $\tau_a$. Counting more carefully, we find the collinear and soft power corrections to the $\tau_a$ distributions to be in powers of $(\Lqcd/(\tau_a Q))^{2-a}$ and $\Lqcd/(\tau_aQ)$, respectively. Then for $a<1$, the soft power corrections are dominant, and we can ignore the collinear power corrections to first approximation. For $a> 1$, the collinear power corrections dominate. For the broadening $B$ they are of the same order.  Furthermore, in this region, there are also other sources of power corrections we have chosen to ignore, such as recoil of the thrust axis.\cite{BKS03,broadening1,broadening2} For these reasons we limit our treatment to $\tau_a$ with $a<1$ as well as other event shapes such as the $C$-parameter for which the soft power corrections dominate.

\section{Soft Power Corrections}

We now turn our attention solely to the power corrections coming from the soft function in the factorized event shape distribution (\ref{convolution}). By studying the energy flow and boost properties of the soft radiation, we will be able to show the universality of the power corrections to the mean values of event shapes. This universality will be violated in shifts of higher moments, but for the angularity distributions, we will be able to extract useful information about the nature of the soft radiation from the details of these violations.

\subsection{Energy Flow}

The delta functions of the event shape variables in the jet and soft functions (\ref{jetfunction}) and (\ref{softfunction}) can be expressed in terms of the \emph{transverse momentum flow operator}, $\mathcal{E}_T(\eta)$, defined by its action on a state $\ket{N}$:
\begin{equation}
\mathcal{E}_T(\eta) \ket{N} = \sum_{i\in N}\abs{\vect{p}_i^\perp}\delta(\eta-\eta_i)\ket{N},
\label{ETdef}
\end{equation}
adding up the magnitudes of transverse momenta of particles with rapidity $\eta$.
These operators are simply related to the energy flow operators\cite{flow} $\mathcal{E}(\hat n)$, which act on states according to:
\begin{equation}
\mathcal{E}(\hat n) \ket{N} = \sum_{i\in N}E_i\delta^2(\hat n - \hat n_i)\ket{N},
\end{equation}
measuring energy flowing in the direction of the unit vector $\hat n$. With $\delta^2(\hat n - \hat n_i) = \delta(\cos\theta-\cos\theta_i)\delta(\phi-\phi_i)$, and $\eta = \ln\cot(\theta/2)$, the two operators are related by:
\begin{equation}
\mathcal{E}_T(\eta) = \frac{1}{\cosh^3\eta}\int_0^{2\pi}\! d\phi\,\mathcal{E}(\hat n).
\end{equation}
The energy flow operator can be represented in terms of the energy-momentum tensor on a sphere at infinity\cite{flowops1,flowops2,flowops3,flowops4}, and is closely related to the energy-energy correlations\cite{BBEL1,BBEL2}. 

Using the operator $\mathcal{E}_T(\eta)$, the soft function can be rewritten:
\begin{equation}
S_e(e) = \frac{1}{N_C}\Tr\sum_{X_u} \bra{0}\bar T[Y_{\bar n} Y_n^\dag]\delta\left(e - \frac{1}{Q}\int_{-\infty}^\infty\!d\eta\,\mathcal{E}_T(\eta)f_e(\eta)\right)\ket{X_u}\bra{X_u}T[Y_n\bar Y_n^\dag]\ket{0}.
\label{softfunctionET}
\end{equation}
Since there is no restriction on the usoft states in the soft function $S_e(e)$ in Eq.~(\ref{softfunctionET}), we can perform the sum over states to obtain
\begin{equation}
S_e(e) = \frac{1}{N_C}\Tr\bra{0}\overline Y_{\bar n}^\dag Y_n^\dag \delta\left(e - \frac{1}{Q}\int_{-\infty}^\infty\! d\eta\,\mathcal{E}_T(\eta)f_e(\eta)\right)Y_n \overline Y_{\bar n}\ket{0}.
\label{softfunctionnosum}
\end{equation}
The time- and anti-time-ordering operators have been removed by using Wilson lines $\overline Y_{n,\bar n}$ in the anti-fundamental representation of $SU(N)$, and the space-like separation $(ns - \bar ns')^2 = -4ss' <0$ of the $n$ and $\bar n$ fields.\cite{BLMW}

\subsection{Boost Invariance and Universality}

Inserting the soft function (\ref{softfunctionnosum}) back into the convolution (\ref{convolution}), we can take moments of the full distribution. The first moment $\langle e\rangle_{\text{PT}}$ of the perturbative distribution $\sigma_J(e_J)$ is shifted by the first moment of the soft function:
\begin{equation}
\langle e\rangle = \langle e\rangle_{\text{PT}} + \delta\langle e\rangle_{\text{NP}},
\end{equation}
where 
\begin{equation}
\delta\langle e\rangle_{\text{NP}} = \frac{1}{Q}  \int_{-\infty}^\infty\! d\eta\,f_e(\eta)\frac{1}{N_C}\Tr \bra{0}\overline Y_{\bar n}^\dag Y_n^\dag\mathcal{E}_T (\eta)Y_n\overline Y_{\bar n}\ket{0}.
\label{shiftmean}
\end{equation}
We can make use of boost properties of the operators in the soft matrix element to simplify this quantity. Insert factors of $1 = U_{\eta'}^{-1}U_{\eta'}$, where $U_{\eta'}$ is the operator implementing a Lorentz boost by a rapidity $\eta'$ along the thrust axis, around each operator in the matrix element:
\begin{equation}
\bra{0}U_{\eta'}^{-1}U_{\eta'} \overline Y_{\bar n}^\dag U_{\eta'}^{-1}U_{\eta'} Y_n^\dag U_{\eta'}^{-1}U_{\eta'}\mathcal{E}_T(\eta)U_{\eta'}^{-1}U_{\eta'}Y_n U_{\eta'}^{-1}U_{\eta'}\overline Y_{\bar n}U_{\eta'}^{-1}U_{\eta'}\ket{0}.
\end{equation}
Now we use the transformation properties of each object in this matrix element:
\begin{equation}
U_{\eta'}\ket{0} = \ket 0, \qquad
U_{\eta'} Y_n U_{\eta'}^{-1} = Y_n, \qquad
U_{\eta'}\mathcal{E}_T(\eta)U_{\eta'}^{-1} = \mathcal{E}_T(\eta + \eta'),
\end{equation}
where the boost invariance of $Y_n$ can be seen from:
\begin{equation}
Y_n = P\exp\left[ig\int_0^\infty ds\,n\cdot A_{us}(ns)\right]\rightarrow P\exp\left[ig\int_0^\infty ds\,\e^{\eta'} n\cdot A_{us}(\e^{\eta'} n s)\right] = Y_n,
\end{equation}
by a simple change of variables $s' = \e^{-\eta'}s$ to give the final equality. This is a manifestation of the reparameterization invariance of SCET.\cite{RPI} The boost transformation of $\mathcal{E}_T(\eta)$ is seen easily from its definition (\ref{ETdef}):
\begin{align}
U_{\eta'}\mathcal{E}_T(\eta)U_{\eta'}^{-1}\ket{N(\{\eta_i\})} &= U_{\eta'}\mathcal{E}_T(\eta)\ket{N(\{\eta_i - \eta'\})} \nonumber \\
& = \sum_{i\in N}U_{\eta'}\abs{\vect{p}_i^\perp}\delta(\eta - \eta_i + \eta')\ket{N(\{\eta_i-\eta'\})} \nonumber \\
&= \mathcal{E}_T(\eta + \eta')\ket{N(\{\eta_i\})}.
\end{align}
This means that the soft matrix element in Eq.~(\ref{shiftmean}) is independent of the rapidity chosen as the argument of $\mathcal{E}_T(\eta)$:
\begin{equation}
\bra{0}\overline Y_{\bar n}^\dag Y_n^\dag\mathcal{E}_T (\eta)Y_n\overline Y_{\bar n}\ket{0} = \bra{0}\overline Y_{\bar n}^\dag Y_n^\dag\mathcal{E}_T (\eta+\eta')Y_n\overline Y_{\bar n}\ket{0} \equiv \mathcal{A}
\end{equation}
This matrix element does not depend on $\eta$ at all, and so is just a constant $\mathcal{A}$. We thus take it out of the integral in Eq.~(\ref{shiftmean}), leaving
\begin{equation}
\langle e\rangle = \langle e\rangle_{\text{PT}} + c_e\frac{\mathcal{A}}{Q},
\end{equation}
where
\begin{equation}
c_e = \int_{-\infty}^\infty\! d\eta\,f_e(\eta),\qquad
\mathcal{A} = \langle\mathcal{E}_T(0)\rangle,
\label{meaning}
\end{equation}
the Wilson lines and color trace being understood in the expectation value of $\mathcal{E}_T(0)$. We have chosen the argument of $\mathcal{E}_T(\eta)$ to be $0$, but could have chosen any rapidity. For the angularities, $c_{\tau_a} = 2/(1-a)$, and for the $C$-parameter, $c_C = 3\pi$.
 
These observations reproduce the prediction (\ref{universalshift}) for the universal shift in the mean value of event shapes. We have not made any assumptions about single or low-order gluon emission in the final state. This result relied rather on the feature of the SCET Lagrangian at leading-order in $\lambda$ that only the $n\cdot A_{us}$ component of usoft gluons couples to $n$-collinear particles ($\bar n\cdot A_{us}$ to $\bar n$-collinear particles), so that these couplings can be removed from the Lagrangian by the redefinition (\ref{Yredef}) of the collinear fields, and on the boost invariance of the usoft Wilson lines.  The observation that the $e$-dependent effects of soft radiation enter  only through the integral in (\ref{meaning}) for $c_e$  was already noted in the derivation of the Milan factor.\cite{Milan} Here, we have obtained in addition a field-theoretic interpretation for the quantity $\mathcal{A}$, and, at leading order in the SCET expansion in $\lambda$, demonstrated its universality exactly to all orders in soft gluon emission.

Through same line of reasoning we discover that the prediction of a universal shift in the full distribution is spoiled by nontrivial correlations in the energy flow of soft radiation. For example, the shift in the second moment of the perturbative distribution due to soft radiation is given by:
\begin{equation}
\delta\langle{e^2}\rangle_{\text{NP}} = \frac{1}{Q^2}\int_{-\infty}^\infty\! d\eta_1\int_{-\infty}^\infty\! d\eta_2\, f_e(\eta_1)f_e(\eta_2)\bra{0}\overline Y_{\bar n}^\dag Y_n^\dag \mathcal{E}_T(\eta_1)\mathcal{E}_T(\eta_2) Y_n\overline Y_{\bar n}\ket{0}.
\end{equation}
Boost properties of the soft operators can be used to eliminate dependence on one of the rapidity variables but not both. Thus this shift does not simplify to $(c_e\mathcal{A}/Q)^2$, as a universal shift (\ref{universalshift}) of the full distribution would require.

\subsection{Power Corrections to Angularities}

More information can be gleaned from the behavior of power corrections to the angularity distributions.
First, we transform the soft function $S_e(e)$ to moment space via the Laplace transform,
\begin{equation}
\tilde S_e(\nu)  = \int_0^\infty \! de\, \e^{-\nu e}S_e(e) = \left\langle\exp\left[-\frac{\nu}{Q}\int_{-\infty}^\infty\!d\eta\,\mathcal{E}_T(\eta)f_e(\eta)\right]\right\rangle,
\end{equation}
which exponentiates in terms of cumulant moments:
\begin{equation}
\tilde S_e(\nu) = \tilde S_e(0)\exp\left[\sum_{n=1}^\infty \frac{1}{n!}\left(-\frac{\nu}{Q}\right)^n\Cumulant{\left(\int_{-\infty}^\infty\! d\eta\,\mathcal{E}_T(\eta)f_e(\eta)\right)^n}\right],
\end{equation}
where the cumulants are defined according to $\cumulant{X} = \langle X\rangle$, $\cumulant{X^2} = \langle X^2\rangle - \langle X\rangle^2$, and so on.

For the angularity shape function, we may compare this form to the prediction (\ref{angularityexponentiated})  based on the behavior of resummed perturbation theory. The parameters $\lambda_n$ in Eq.~(\ref{angularityexponentiated}) correspond here to the cumulants of energy flow operators:
\begin{equation}
\frac{1}{1-a}\lambda_n(a) = \Cumulant{\left(\int_{-\infty}^\infty\! d\eta\,\mathcal{E}_T(\eta)\e^{-\abs{\eta}(1-a)}\right)^n},
\end{equation}
where we now allow a possible $a$ dependence in the coefficients $\lambda_n$. For $n=1$, we find simply
\begin{equation}
\frac{1}{1-a}\lambda_1 = \frac{2}{1-a}\mathcal{A},
\end{equation}
and so $\lambda_1$ is truly $a$-independent. However, beginning with $n=2$, the coefficients $\lambda_n$ depend, in general, on $a$. For example,
\begin{equation}
\frac{1}{1-a}\lambda_2(a) = \frac{1}{2}\int_{-\infty}^\infty \! d\eta_1\int_{-\infty}^\infty\! d\eta_2\,\e^{-(1-a)(\abs{\eta_1} + \abs{\eta_2})}\cumulant{\mathcal{E}_T(\eta_1)\mathcal{E}_T(\eta_2)}.
\end{equation}
We again insert Lorentz boosts into the matrix element, and this time are able to eliminate dependence on one but not both rapidities, and perform one integration:
\begin{equation}
\frac{1}{1-a}\lambda_2(a) = \frac{1}{2(1-a)}\int_{-\infty}^\infty\! d\eta\,[1 + (1-a)\abs{\eta}]\e^{-\abs{\eta}(1-a)}\cumulant{\mathcal{E}_T(0)\mathcal{E}_T(\eta)},
\label{2pointintegral}
\end{equation}
so correlator still depends on the rapidity gap between the two insertions of $\mathcal{E}_T$. 

If the two-point correlations of energy flow $\cumulant{\mathcal{E}_T(0)\mathcal{E}_T(\eta)}$ are negligible for rapidities $\abs{\eta}\gg\frac{1}{1-a}$, then the coefficient of the matrix element in (\ref{2pointintegral}) can be expanded about $\eta=0$,
\begin{equation}
\lambda_2(a) = \frac{1}{2}\int_{-\infty}^\infty\! d\eta\,\left\{1 - \frac{1}{2}[(1-a)\eta]^2 + \frac{1}{3}[(1-a)\abs{\eta}]^3+\cdots\right\}\cumulant{\mathcal{E}_T(0)\mathcal{E}_T(\eta)}.
\end{equation}
Thus the $a$-independence of $\lambda_2(a)$ is tied to the assumption of negligible correlations in energy flow of soft radiation.\cite{scaling1}

The violation of the scaling rule (\ref{scalingrule}) can itself be used to deduce information about the correlations in the final state soft radiation. The quantity,
\begin{equation}
\mathcal{C}_2(a) = \frac{1}{2}\int_{-\infty}^\infty \! d\eta\,\e^{-\abs{\eta}(1-a)}\cumulant{\mathcal{E}_T(0)\mathcal{E}_T(\eta)},
\end{equation}
is essentially a Laplace transform of the two-point correlator to the variable $a$. Equation~(\ref{2pointintegral}) relates $\mathcal{C}_2(a)$ to the quantity $\lambda_2(a)$:
\begin{equation}
\mathcal{C}_2(a) - \frac{\partial}{\partial\ln(1-a)}\mathcal{C}_2(a)  = \lambda_2(a),
\end{equation}
whose solution is
\begin{equation}
\mathcal{C}_2(a) = (1-a)\int_{-\infty}^a\frac{da'}{(1-a')^2}\lambda_2(a'),
\end{equation}
assuming the correlator vanishes for $a\rightarrow-\infty$.
Thus, the observation of $a$-dependence in the coefficient $\lambda_2(a)$ in the expansion of the angularity shape function (\ref{angularityexponentiated}) yields direct information on correlations of the energy flow of soft radiation at separated rapidities. This information can, in principle, be extracted from data in $e^+e^-$ collisions at LEP.

\section{Conclusions}

Event shapes provide a useful laboratory for the study of perturbative and nonperturbative properties of hadronic jets. The continuous adjustable parameter $a$ in the angularities provides an extra source of information about the nature of the hadronization process in jet formation. Soft-collinear effective theory provides an organized framework in which to understand factorization, running, and power corrections that have traditionally been described in the language of full QCD. In particular, the decoupling of collinear and usoft modes in the leading-order Lagrangian facilitates factorization, and the form of the soft function and its boost properties lead to a proof of the universality of power corrections to the mean values of event shapes, as well as the extraction of useful information from the violation of universality or scaling of power corrections to higher moments. SCET systematizes the approximations that lead to useful predictions, and sets up the framework that would allow one to calculate corrections, such as those suppressed by $\lambda$ or that come from including operators with more jets in the matching from QCD. A fuller understanding of power corrections to event shapes in $e^+e^-$ collisions helps point the way to the eventual understanding of jets produced in hadron collisions and the nonperturbative dynamics of QCD.

\section*{Acknowledgments}

I would like to thank C. Bauer, A. Manohar, G. Sterman, and M. Wise for their collaboration on the work reviewed here, and E. Gardi for comments on the manuscript. This work was supported in part by the U.S. Department of Energy under Contract No. DE-FG02-00ER41132.

\end{document}